\documentclass{amsart} 
\usepackage{amssymb}
 \usepackage{a4}
 \begin{document}
 \newtheorem{theorem}{Theorem}[section]
 \newtheorem{lemma}[theorem]{Lemma}
 \newtheorem{remark}[theorem]{Remark}
 \newtheorem{definition}[theorem]{Definition}
 \newtheorem{corollary}[theorem]{Corollary}
 \def\deg{\operatorname{deg}}
 \def\pext{\operatorname{ext}}
 \def\pint{\operatorname{int}}
 \def\tr{{\operatorname{tr}}}
 \def\id{{\operatorname{id}}}
 \def\Pspan{{\operatorname{Span}}}
 \def\BB{{\mathcal{B}}}
 \def\Tr{{\operatorname{Tr}}}
 \def\Span{{\operatorname{Span}}}
 \font\pbglie=eufm10
 \def\rr{{\text{\pbglie R}}}
 \def\xxe{{\text{\pbglie e}}}
 \def\xxi{{\text{\pbglie i}}}
 \def\cc{b}
 \def\ee{\Theta_{m+2,m}}
 \def\EE{\Xi_{m+2,m}}
 \def\EX{\mathcal{E}}
 \def\qedbox{\hbox{$\rlap{$\sqcap$}\sqcup$}}
 \makeatletter
 \renewcommand{\theequation}{%
 \thesection.\alph{equation}}
 \@addtoreset{equation}{section}
 \makeatother
\title[Invariance Theory]
{Supertrace divergence terms for the Witten Laplacian} 
\author[Gilkey, Kirsten, and Vassilevich]{P. Gilkey, K. Kirsten, and D.
Vassilevich}
\begin{address}{PG: Mathematics Department, University of Oregon, 
Eugene Or 97403 USA}\end{address}
\begin{email}{gilkey@darkwing.uoregon.edu}\end{email}
\begin{address}{KK: Department of Mathematics,
Baylor University, Waco, TX 76798 USA 
and Max-Planck-Institute for Mathematics in the Sciences, 
Inselstrasse 22-26, 04103 Leipzig Germany}\end{address}
\begin{email}{klaus.kirsten@mis.mpg.de and 
Klaus\_Kirsten@baylor.edu}\end{email}
\begin{address}{DV: Max-Planck-Institute for 
Mathematics in the Sciences, Inselstrasse 22-26, 04103 Leipzig Germany}
\end{address}
\begin{email}{vassil@itp.uni-leipzig.de}
\end{email}

 \begin{abstract} We use invariance theory to compute the divergence 
 term $a_{m+2,m}^{d+\delta}$ in the super trace for
 the twisted de Rham complex for a closed Riemannian 
 manifold.\end{abstract}
\maketitle
 \section{Introduction}

 Let $(M,g)$ be a compact $m$ dimensional Riemannian manifold without 
 boundary. The fundamental solution of
 the heat equation $e^{-tD}$ for an operator of Laplace type $D$ on $M$ 
 is
 an infinitely smoothing operator. Let $f\in
 C^\infty(M)$ be an auxiliary smooth `smearing' function. Work of Seeley 
 \cite{Se68} shows
 the smeared heat trace
 has a complete asymptotic expansion as
 $t\downarrow0$ of the form:
 $$\Tr_{L^2}(fe^{-tD})\sim\textstyle\sum_{n\ge0}a_{n,m}(f,D)t^{(n-m)/2}.$$
 The {\it heat trace invariants} $a_{n,m}$ vanish if $n$ is odd; if $n$ 
 is even, there are local invariants
 $a_{n,m}(x,D)$ so that
 $$a_{n,m}(f,D)=\textstyle\int_Mf(x)a_{n,m}(x,D)\operatorname{dvol}_g(x).$$
 The function $f$ localizes the problem and permits us to recover 
 divergence terms
 which would otherwise not be detected.

 Let $\phi$ be an auxiliary smooth function called the dilaton. We twist 
 the exterior derivative $d$ and the
 coderivative $\delta_g$ to define
 $$d_\phi:=e^{-\phi}de^\phi\quad\text{and}\quad\delta_{\phi,g}:=e^{\phi}\delta_ge^{-\phi}.$$
 We denote the associated Laplacian by $\Delta_{\phi,g}^p$ on 
 $C^\infty(\Lambda^p(M))$. It appears in supersymmetric
 quantum mechanics \cite{ABI} and in the study of Morse theory 
 \cite{Wit82}. It also is used to study quantum
 $p$ form fields interacting with a background dilaton 
 \cite{GKVZ02,VZ00}. 

 Let $\chi(M):=\textstyle\sum_p(-1)^p\dim H^p(M;\mathbb{R})$ be the 
 Euler-Poincar\'e
 characteristic of $M$. Arguments of McKean and Singer \cite{McSi67} 
 extend to the twisted setting to show
 \begin{equation}\label{eqn1.a}\textstyle\sum_p(-1)^p\Tr_{L^2}(e^{-t\Delta_{\phi,g}^p})=\chi(M).\end{equation}
 We define the local supertrace asymptotics by setting:
 $$
 a_{n,m}^{d+\delta}(\phi,g)(x):=\textstyle\sum_p(-1)^pa_{n,m}(x,\Delta_{\phi,g}^p).
 $$ 
 We expand the left hand side of equation (\ref{eqn1.a}) and then equate 
 powers of $t$ to see:
 \begin{equation}\label{eqn1.b}\textstyle\int_Ma_{n,m}^{d+\delta}(\phi,g)(x)
 \operatorname{dvol}_g(x)=\left\{
 \begin{array}{lll}
 \chi(M)&\text{if}&n=m,\\
 0&\text{if}&n\ne m.
 \end{array}\right.\end{equation}

 Let $R_{ijkl}$ be the components of the Riemann curvature tensor 
 relative to a local
 orthonormal frame for the tangent bundle with the sign convention that 
 $R_{1221}=+1$ on
 the unit sphere $S^2\subset\mathbb{R}^3$. 
We adopt the Einstein convention and sum over repeated indices. If $I=(i_1,...,i_m)$ and 
 $J=(j_1,...,j_m)$ are $m$ tuples of indices, let
 $$\varepsilon_J^I:=
 g(e_{i_1}\wedge...\wedge e_{i_m},e_{j_1}\wedge...\wedge e_{j_m})$$
 be the totally anti-symmetric tensor and let
 $$\mathcal{R}_{J,s}^{I,t}:=R_{i_si_{s+1}j_{s+1}j_s}...R_{i_{t-1}i_tj_tj_{t-1}};$$
we set
 $\mathcal{R}_{J,s}^{I,t}=1$ for $t<s$.

 \begin{theorem}\label{thm1.1}\begin{enumerate}
 \item If $n<m$ or if $n$ is odd, then 
 $a_{n,m}^{d+\delta}(\phi,g)(x)=0$.
 \item If $m=2\bar m$, then $a_{m,m}^{d+\delta}(\phi,g)(x)
 =\textstyle\frac{1}{8^{\bar m}\pi^{\bar m}\bar
 m!}\varepsilon_I^J
 \mathcal{R}_{J,1}^{I,m}$.
 \item If $m=2\bar m+1$, then 
 $a_{m+1,m}^{d+\delta}(\phi,g)(x)=\textstyle\frac1{\sqrt{\pi}8^{\bar m}\pi^{\bar m}
 \bar m!}\varepsilon_J^I\phi_{;i_1j_1}\mathcal{R}_{J,2}^{I,m}$.
 \end{enumerate}
 \end{theorem}
 
 Assertions (1) and (2) were proved in the untwisted case ($\phi=0$) by 
 Atiyah, Bott, and Patodi \cite{ABP73}, by Gilkey \cite{PG73}, and by 
 Patodi \cite{Pa70}. This provided a heat
 equation proof of the classical Chern-Gauss-Bonnet \cite{C44} theorem. 
 Assertions (1) and (2) in the twisted setting were
 established in
 \cite{GKVZ02} and the divergence term $a_{m+1,m}^{d+\delta}$ was 
 identified in \cite{GKV02a}. Our previous paper
 \cite{GKV02a} dealt with the odd dimensional case for manifolds with 
 boundary. The present paper computes
 $a_{m+2,m}^{d+\delta}$ for closed even dimensional manifolds. This 
 requires significantly different techniques. We note
 that some information concerning $a_{m+2,m}^{d+\delta}$ was derived 
 earlier
 in \cite{PG79} using an entirely different approach.

 The main new result of this paper is the following:

 \begin{theorem}\label{thm1.2} Let $M$ be a closed Riemannian manifold 
 of dimension $m=2\bar m$.
 \begin{eqnarray*}
 a_{m+2,m}^{d+\delta}&=&\textstyle
 \frac{1}{\pi^{\bar m}8^{\bar m}\bar m!}\varepsilon_J^I
 \{4\bar m(\phi_{;i_1j_1}\phi_{;i_2}\mathcal{R}_{J,3}^{I,m})_{;j_2}
 +\textstyle\frac1{12}(\mathcal{R}_{J,1}^{I,m})_{;kk}\\
 &+&\textstyle\frac{\bar
 m}6(R_{i_1i_2kj_1;k}\mathcal{R}_{J,3}^{I,m})_{;j_2}\}.\end{eqnarray*}
 \end{theorem}

 In Section \ref{Sect2}, we recall combinatorial formulas for the 
 invariants $a_{n,m}(x,D)$ for $n=0,2,4,6$ \cite{PG94};
 formulas for $a_8$, and for $a_{10}$ are available 
 \cite{AmBeOc89,Av90,Ven98}. These formulas become very
 complicated as $n$ increases and it seems hopeless to try to establish 
 Theorem \ref{thm1.2} via direct computation even
 for $m=6$ and
 $m=8$. There are also closed formulas available due to Polterovich
 \cite{Pol00}. However, it does not seem possible to make a direct use 
 of these formulas to derive Theorem
 \ref{thm1.2}. Instead, we proceed indirectly. In Section \ref{Sect3}, 
 we establish some functorial properties
 of these invariants and use invariance theory to prove the following 
 result:

 \begin{lemma}\label{lem1.3} If $m$ is even, then there exist universal 
 constants so that
 \begin{eqnarray*}
 a_{m+2,m}^{d+\delta}(\phi,g)&=&c_{m+2,m}^1(\varepsilon_J^I\phi_{;i_1j_1}\phi_{;i_2}\mathcal{R}_{J,3}^{I,m})_{;j_2}
 +c_{m+2,m}^2(\varepsilon_J^I\mathcal{R}_{J,1}^{I,m})_{;kk}
 \\&&
 +c_{m+2,m}^3(\varepsilon_J^IR_{i_1i_2kj_1;k}\mathcal{R}_{J,3}^{I,m})_{;j_2}
 .\end{eqnarray*}
 \end{lemma} 

 We shall complete the proof of Theorem \ref{thm1.2} in Section 
 \ref{Sect4} by evaluating these normalizing
 constants. The new features of this investigation are that both H. 
 Weyl's first and second main theorems of invariance
 theory
 \cite{We46} play a crucial role as does the analysis of the formal 
 cohomology groups of spaces of $p$ form valued
 invariants
 \cite{PG75}. Thus we expect that the techniques presented in this paper 
 will be useful in other similar investigations
 of this type.

 \section{Local Formulae for the heat trace invariants}\label{Sect2}

 If $D$ is an operator of Laplace type, then there is a canonical 
 connection
 $\nabla$ on the underlying vector bundle that, together with the 
 Levi-Civita connection, we use to
 covariantly differentiate tensors of all types - we denote multiple 
 covariant differentiation by `;'.
 There is also a canonical endomorphism
 $E$ so that 
 $$D=-(\Tr(\nabla^2)+E)\quad\text{i.e.}\quad Du=-(u_{;ii}+Eu).$$
 Let $\Omega_{ij}$ be the curvature of the connection $\nabla$. Let
 $\rho_{ij}=R_{ikkj}$ be the Ricci tensor, and let $\tau=\rho_{jj}$.

 The heat trace invariants $a_{n,m}$ for an operator of Laplace type can 
 be expressed in this formalism \cite{PG94}:
 \begin{theorem}\label{thm2.1} \ \begin{enumerate}
 \smallskip\item $a_{0,m}=(4\pi)^{-m/2}\Tr\{\operatorname{Id}\}$.
 \smallskip\item 
 $a_{2,m}=(4\pi)^{-m/2}\frac16\Tr\{6E+\tau\operatorname{Id}\}$.
 \smallskip\item $a_{4,m}=(4\pi)^{-m/2}\frac1{360}\Tr\{
 60E_{;kk}+60\tau 
 E+180E^{2}+(12\tau_{;kk}+5\tau^{2}$\smallbreak\qquad
 $-2|\rho|^{2}+2|R|^{2})\operatorname{Id}+30\Omega_{ij}\Omega_{ij}\}$.
 \smallskip\item $a_{6,m}=(4\pi)^{-m/2}\Tr\{(\frac{18}{7!}\tau_{;iijj}
 +\frac{17}{{7!}}\tau_{;k}\tau_{;k}-\frac2{7!}\rho_{ij;k}\rho_{ij;k}
 -\frac4{7!}\rho_{jk;n}\rho_{jn;k}$
 \smallbreak\qquad
 $+\frac9{7!}R_{ijkl;n}R_{ijkl;n}+\frac{28}{7!}\tau\tau_{;nn}
 -\frac8{7!}\rho_{jk}\rho_{jk;nn}+\frac{24}{7!}\rho_{jk}\rho_{jn;kn}$
 \smallbreak\qquad
 $+\frac{12}{7!}R_{ijkl}R_{ijkl;nn}+
 \frac{35}{9\cdot {7!}}\tau^{3}-\frac{14}{3\cdot
 {7!}}\tau\rho^{2}+\frac{14}{3\cdot {7!}}\tau R^{2}
 -\frac{208}{9\cdot {7!}}\rho_{jk}\rho_{jn}\rho_{kn}$
 \smallbreak\qquad
 $-\frac{64}{3\cdot{7!}}\rho_{ij}\rho_{kl}R_{ikjl}
 -\frac{16}{3\cdot {7!}}\rho_{jk}R_{jnli}R_{knli}-
 \frac{44}{9\cdot {7!}}R_{ijkn}R_{ijlp}R_{knlp}$
 \smallbreak\qquad
 $ -\frac{80}{9\cdot {7!}}R_{ijkn}R_{ilkp}R_{jlnp})\operatorname{Id}
  +\frac1{45}\Omega_{ij;k}\Omega_{ij;k}+\frac1{180}\Omega_{ij;j}\Omega_{ik;k}
 +\frac{1}{60}\Omega_{ij;kk}\Omega_{ij}$
 \smallbreak\qquad$
 +\frac1{60}\Omega_{ij}\Omega_{ij;kk}
 -\frac1{30}\Omega_{ij}\Omega_{jk}\Omega_{ki}-\frac1{60}
 R_{ijkn}\Omega_{ij}\Omega_{kn}
 -\frac1{90}\rho_{jk}\Omega_{jn}\Omega_{kn}$
 \smallbreak\qquad
 $+\frac1{72}\tau\Omega_{kn}\Omega_{kn}+\frac1{60}E_{;iijj}+\frac1{6}EE_{;ii}
 +\frac1{12}E_{;i}E_{;i}+\frac16 E^{3}$
 \smallbreak\qquad
 $+\frac1{12}E\Omega_{ij}\Omega_{ij}+\frac1{36}\tau 
 E_{;kk}+\frac1{90}\rho_{jk}E_{;jk}$
 $+\frac1{30}\tau_{;k}E_{;k}+\frac1{12}EE\tau$\smallbreak\qquad$
 +\frac1{30}E\tau_{;kk}
 +\frac1{72}E\tau^{2}-\frac1{180}E|\rho|^2+\frac1{180}E|R|^{2}\}$.\end{enumerate}\end{theorem}


 We refer to \cite{GKV02a} for the proof of the following results:


 \begin{lemma}\label{lem2.2}On the circle, 
 $a_{2,1}^{d+\delta}=\frac1{\sqrt\pi}\phi_{;11}$.\end{lemma}

 \begin{lemma}\label{lem2.3} Let $M=(M_1\times 
 M_2,\phi_1+\phi_2,g_1+g_2)$ decouple as a product. Then
 $a_{n,m}^{d+\delta}(\phi,g)(x_1,x_2)=
 \textstyle\sum_{n_1+n_2=n, n_1\ge m_1, n_2\ge m_2}
 a_{n_1,m_1}^{d+\delta}(\phi_1,g_1)(x_1)\cdot
 a_{n_2,m_2}^{d+\delta}(\phi_2,g_2)(x_2)$.
 \end{lemma}

 \section{Spaces of Invariants}\label{Sect3}

 Let $\mathcal{Q}_m$ be the space of $O(m)$ invariant polynomials in the 
 components of the tensors $\{R,\nabla
 R,\nabla^2R,...,\phi,\nabla\phi,\nabla^2\phi,...\}$.
 Define a grading on
 $\mathcal{Q}_m$ by setting:
 $$\operatorname{weight}(R_{ijkl;\beta})=|\beta|+2\text{ and 
 }\operatorname{weight}(\phi_{;\beta})=|\beta|.$$
 An element $Q\in\mathcal{Q}_m$ is homogeneous of weight
 $n$ if and only if
 $$Q(\phi,c^{-2}g)=c^nQ(\phi,g).$$
 Let
 $\mathcal{Q}_{n,m}\subset\mathcal{Q}_m$ be the set of all $O(m)$ 
 invariant polynomials which are
 homogeneous of weight $n$; we then have a direct sum decomposition:
 $$\mathcal{Q}_m=\oplus_n\mathcal{Q}_{n,m}.$$
 We may use the $\mathbb{Z}_2$ action $\phi\rightarrow-\phi$ to 
 decompose
 $\mathcal{Q}_{n,m}=\mathcal{Q}_{n,m}^+\oplus\mathcal{Q}_{n,m}^-$ where
 $$\mathcal{Q}_{n,m}^\pm:=\{Q\in\mathcal{Q}_{n,m}:Q(-\phi,g)=\pm 
 Q(\phi,g)\}.$$

 The following natural restriction map
 $r:\mathcal{Q}_{n,m}\rightarrow\mathcal{Q}_{n,m-1}$
 will play a crucial role. If $(N,g_N,\phi_N)$ are
 structures in dimension $m-1$, then we can define
 corresponding structures in dimension $m$ by setting 
 $$(M,\phi_M,g_M):=(N\times
 S^1,\phi_N,g_N+d\theta^2)\,.$$
 If
 $x\in N$ is the point of evaluation, we take the corresponding point 
 $(x,1)\in M$ for
 evaluation; which point on the circle chosen is, of course, irrelevant 
 as $S^1$ has a
 rotational symmetry. The restriction map is characterized dually by the 
 formula:
 $$r(Q)(\phi_N,g_N)(x)=Q(\phi_N,g_N+d\theta^2)(x,1)\,.$$
 We can also describe the restriction map $r$ in classical terms. H. 
 Weyl's \cite{We46} first
 theorem of invariance theory implies
 orthogonal invariants are built by contracting indices in pairs, where 
 the indices range from $1$ through $m$. If $P$
 is given in terms of such a Weyl spanning set, then
 $r(P)$ is given in terms of the same Weyl spanning set by restricting 
 the range of summation to be
 from $1$ through
 $m-1$. Thus necessarily $r$ is surjective. We refer to \cite{GKV02a} 
 for the proof of:

 \begin{lemma}\label{lem3.1} If $m$ is even, then
 $a_{m+2,m}^{d+\delta}\in\mathcal{Q}_{n,m}^+\cap\ker(r)$.
 \end{lemma}

 We use H. Weyl's second theorem to see 
 $\mathcal{Q}_{m+2,m}^+\cap\ker(r)$ is generated by
 invariants where we contract $2m$ indices using the $\varepsilon$ 
 tensor and contract the remaining indices in
 pairs, we refer to the discussion in \cite{GKV02a} for details. A 
 direct calculation shows, after some additional
 work to eliminate dependencies, that:
 \begin{eqnarray*} 
 &&\mathcal{Q}_{m+2,m}^+\cap\ker(r)=
 \Span\{\varepsilon_J^I\phi_{;i_1j_1}\phi_{;i_2j_2}\mathcal{R}_{J,3}^{I,m},\ 
 \varepsilon^I_J\phi_{;k}\phi_{;k}
 \mathcal{R}_{J,1}^{I,m},\ 
 \varepsilon^I_J
 R_{k\ell\ell k}\mathcal{R}_{J,1}^{I,m},\\
 &&\quad\varepsilon^I_J\phi_{;i_1}\phi_{;j_1}R_{ki_2j_2k}
 \mathcal{R}_{J,3}^{I,m},\ 
 \varepsilon^I_J
 R_{i_1i_2j_2j_1;kk}\mathcal{R}_{J,3}^{I,m},\ 
 \varepsilon^I_J
 R_{i_1i_2j_2j_1;k}R_{i_3i_4j_4j_3;k}\mathcal{R}_{J,5}^{I,m},\\
 &&\quad
 \varepsilon^I_J
 R_{ki_1j_1k}R_{\ell i_2j_2\ell}\mathcal{R}_{J,3}^{I,m},\ 
 \varepsilon^I_J
 R_{k\ell j_2j_1}R_{i_1i_2k\ell}\mathcal{R}_{J,3}^{I,m},\ 
 \varepsilon_J^IR_{ki_1j_1\ell}R_{ki_2j_2\ell}\mathcal{R}_{J,3}^{I,m},\\
 &&\quad\varepsilon^I_J
 R_{ki_1j_2j_1}R_{kj_3i_3i_2}R_{\ell i_4j_5j_4}R_{\ell j_6i_6i_5}
 \mathcal{R}_{J,7}^{I,m}\}.
 \end{eqnarray*}
 Although this is some gain in simplifying the question, the list of 
 invariants is still quite long. We shall use
 equation (\ref{eqn1.b}) to further reduce the number of invariants to 
 be considered and complete the proof of Lemma
 \ref{lem1.3}.

 Let $\mathcal{Q}_{n,m}^{p,+}$ be the space of $p$ form valued 
 invariants in
 the curvature tensor, the covariant derivatives of the curvature 
 tensor, and the covariant derivatives of $\phi$
 which are even in $\phi$. The exterior co-derivative $\delta_g$ induces 
 a natural map 
 $$\delta_g:\mathcal{Q}_{n,m}^{p,+}\rightarrow\mathcal{Q}_{n+1,m}^{p-1,+}.$$

 Let $i:N\rightarrow N\times\{1\}\subset N\times S^1$. The analysis of 
 \cite{ABP73} shows that $p$ form valued invariants
 are constructed by alternating $p$ indices and by contracting the 
 remaining indices in pairs. The restriction map
 $$r:\mathcal{Q}_{n,m}^p\rightarrow\mathcal{Q}_{n,m-1}^p$$ 
 is defined by restricting the range of summation of the indices 
 involved; it is characterized by the identity:
 $$r(Q)(\phi_N,g_N)=i^*Q(\phi_N,g_N+d\theta^2).$$
 One verifies that $r$ is surjective and that
 $$r\circ\delta_{g_M}=\delta_{g_N}\circ r\,.$$
 The analysis of the
 formal cohomology groups of the spaces of invariants of Riemannian 
 manifolds, which was given in
 \cite{PG75}, then extends immediately to this more general setting to 
 yield:

 \begin{lemma}\label{lem3.2}\begin{enumerate}
 \item If $Q\in\mathcal{Q}_{n,m}^+$, if $\int_MQ(\phi,g)=0$ for all 
 $(\phi,g)$, and if $n\ne m$, then there exists
 $Q^1\in\mathcal{Q}_{n-1,m}^{1,+}$ so that $\delta_gQ^1=Q$.
 \item If $Q^1\in\mathcal{Q}_{n,m}^{1,+}$, 
 if $\delta_gQ^1=0$, and if $n\ne m-1$, then there exists
 $Q^2\in\mathcal{Q}_{n-1,m}^{2,+}$ so that $Q^1=\delta_gQ^2$.
 \end{enumerate}
 \end{lemma}

 The first assertion shows that any scalar invariant which always 
 integrates to zero is canonically a divergence and
 that any $1$ form valued invariant which is co-closed is canonically 
 co-exact. The restriction on the weight is a
 technical one which plays no role as we shall take $n=m+2$ in 
 assertion (1) and $n=m+1$ in assertion (2).

 We use this result to show
 \begin{lemma}\label{lem3.3} If $m$ is even, then there exists a $1$ 
 form valued invariant
 $Q_{m+1,m}^1$ in $\mathcal{Q}_{m+1,m}^{1,+}\cap\ker(r)$ so that 
 $\delta_gQ_{m+1,m}^1=a_{m+2,m}^{d+\delta}$. 
 \end{lemma}

 \begin{proof} By equation (\ref{eqn1.b}) and Lemma \ref{lem3.2} (1),
 there exists $\bar Q_{m+1,m}^1\in\mathcal{Q}_{m+1,m}^{1,+}$ so
 \begin{equation}\label{eqn3.x}\delta_g\bar 
 Q_{m+1,m}^1(\phi,g)=a_{m+2,m}^{d+\delta}(\phi,g).\end{equation}
 Unfortunately, $r(\bar Q_{m+1,m}^1)$ need not be zero and we must 
 correct for this.
 Since $r(a_{m+2,m}^{d+\delta})=0$,
 $$0=r(\delta_g\bar Q_{m+1,m}^1)=\delta_gr(\bar Q_{m+1,m}^1).$$
 Thus by Lemma \ref{lem3.2} (2), there
 exists $\bar Q_{m,m-1}^2\in\mathcal{Q}_{m,m-1}^2$ so that
 $$
 r(\bar Q_{m+1,m}^1)=\delta_g(\bar Q_{m,m-1}^2).
 $$
 We observed above that $r$ is surjective. Thus we can find 
 $Q_{m,m}^2\in\mathcal{Q}_{m,m}^{2,+}$ so that:
 $$r(Q_{m,m}^2)=\bar Q_{m,m-1}^2.$$
 We complete the proof by setting $Q_{m+1,m}^1:=\bar 
 Q_{m+1,m}^1-\delta_g(Q_{m,m}^2)$ and computing:
 \medbreak\quad\qquad
 $\delta_g(Q_{m+1,m}^1)=\delta_g(\bar 
 Q_{m+1,m}^1)-\delta_g^2(Q_{m,m}^2)=\delta_g(\bar Q_{m+1,m}^1)=a_{m+2,m}^{d+\delta}$,
 \medbreak\quad\qquad
 $r(Q_{m+1,m}^1)=r(\bar Q_{m+1,m}^1)-r(\delta_g(Q_{m,m}^2))
 =r(\bar Q_{m+1,m}^1)-\delta_g(r(Q_{m,m}^2))$
 \medbreak\quad\qquad\quad $
 =r(\bar Q_{m+1,m}^1)-\delta_g(\bar Q_{m,m-1}^2)=0$.
 \end{proof}

 For $m$ even, we define elements of
 $\mathcal{Q}_{m+1,m}^{1,+}\cap\ker(r)$ by setting:
 $$\begin{array}{ll}
 \Xi_{m+1,m}^{1,\ell}:=\varepsilon_J^I\phi^\ell\phi_{;i_1j_1}\phi_{;i_2}\mathcal{R}_{J,3}^{I,m}e^{j_2}
 &(\ell\text{ even}),\\
 \Xi_{m+1,m}^{2,\ell}:=\varepsilon_J^I\phi^\ell 
 R_{i_1i_2j_2j_1;k}\mathcal{R}_{J,3}^{I,m}e^k
 &(\ell\text{ even}),\\
 \Xi_{m+1,m}^{3,\ell}:=\varepsilon_J^I\phi^\ell 
 R_{i_1i_2kj_1;k}\mathcal{R}_{J,3}^{I,m}e^{j_2}
 &(\ell\text{ even}),\\
 \Xi_{m+1,m}^{4,\ell}:=\varepsilon_J^I\phi^\ell\phi_{;k}\mathcal{R}_{J,1}^{I,m}e^k,&(\ell\text{ 
 odd}),\\
 \Xi_{m+1,m}^{5,\ell}:=\varepsilon_J^I\phi^\ell\phi_{;i_1}R_{i_2kkj_2}\mathcal{R}_{J,3}^{I,m}e^{j_1}
 &(\ell\text{ odd}).
 \end{array}$$

 \begin{lemma}\label{lem3.4} If $m=2\bar m$ is even, then
 $\{\mathcal{Q}_{m+1,m}^{1,+}\cap\ker(r)\}=\Span\{\Xi_{m+1,m}^{i,\ell}\}_{i,\ell}$.
 \end{lemma}

 \begin{proof} We use H. Weyl's theorem on the invariants of the 
 orthogonal group. Let
 $$A=\phi^\ell\phi_{;\alpha_1}...\phi_{;\alpha_u}R_{i_1j_1k_1l_1;\beta_1}...R_{i_vj_vk_vl_v;\beta_v}e^h$$
 be a typical $1$ form valued monomial 
 where $|\alpha_\nu|\ge1$ and $\ell+u$ is even. Note that:
 $$n=\textstyle\sum_\mu |\alpha_\mu | + \sum_\nu (|\beta_\nu |+2).$$ 
 We must contract
 $2m$ indices using the $\varepsilon$ tensor and contract the remaining 
 indices in pairs; we refer to \cite{GKV02a}
 where this was discussed in some detail for scalar invariants -- the 
 extension to $1$ form valued invariants is
 similar. We may estimate:
 \begin{eqnarray}
 \label{eqn3.a}2m&\le&\text{ number of indices in }A\\
 &=&\textstyle\sum_\mu|\alpha_\mu|+\sum_\nu(|\beta_\nu|+4)+1=n+2v+1\nonumber\\
 \label{eqn3.b}&=&2n+1-\textstyle\sum_\mu|\alpha_\mu|-\sum_\nu|\beta_\nu|\le 
 2n+1
 \end{eqnarray}
 We set $n=m+1$. Since $2m$ and $m+1+2v+1$ are both even, 
 the inequality in equation (\ref{eqn3.b}) must be
 strict and represents an increase either of $1$ or of $3$.

 \smallbreak Suppose first that equation (\ref{eqn3.a}) is an equality. 
 Then all the $2m$ indices present in $A$ are
 contracted using the
 $\varepsilon$ tensor. We can commute covariant derivatives at the cost 
 of introducing additional curvature terms. Thus
 since all indices are to be contracted using the $\varepsilon$ tensor, 
 we may assume $|\alpha_\mu|\le2$ for all $\mu$.
 Furthermore, by the first and second Bianchi identity, at most
 $2$ indices can be alternated in
 $R_{ijkl;\beta}$. Thus $|\beta_\nu|=0$ for all $\nu$ so 
 $\textstyle\sum_\mu|\alpha_\mu|=3$. This leads to the invariants
 $\Xi_{m+1,m}^{1,\ell}$.

 \smallbreak Suppose next that equation (\ref{eqn3.a}) is not an 
 equality. Then there are $2m+2$ indices and one
 explicit covariant derivative present in $A$; $2m$ indices are 
 contracted using the $\varepsilon$ tensor and two indices
 are contracted as a pair. This yields the invariants 
 $\Xi_{m+1,m}^{i,\ell}$ for $i=2,3,4,5$ and the additional
 invariants:
 $$\begin{array}{ll}
 \Theta_{m+1,m}^{1,\ell}:=\varepsilon_J^I\phi^\ell\phi_{;k}R_{i_1i_2j_2k}\mathcal{R}_{J,3}^{I,m}e^{j_1}
 &(\ell\text{ odd}),\\
 \Theta_{m+1,m}^{2,\ell}:=\varepsilon_J^I\phi^\ell\phi_{;i_1}R_{i_2kj_2j_1}\mathcal{R}_{J,3}^{I,m}e^k
 &(\ell\text{ odd}),\\
 \Theta_{m+1,m}^{3,\ell}:=\varepsilon_J^I\phi^\ell\phi_{;i_1}R_{i_2kj_3j_2}R_{i_3i_4j_4k}\mathcal{R}_{J,5}^{I,m}e^{j_1}
 &(\ell\text{ odd}),\\
 \Theta_{m+1,m}^{4,\ell}:=\varepsilon_J^I\phi^\ell 
 R_{i_1i_2kj_2}R_{i_3i_4j_4j_3;k}\mathcal{R}_{J,5}^{I,m}e^{j_1}
 &(\ell\text{ even}).
 \end{array}$$

 To complete the proof, we must show the invariants 
 $\Theta_{m+1,m}^{i,\ell}$ play no role. Let $U$ and $V$ be collections
 of
 $m+1$ indices. Since $\varepsilon_V^U=0$, we have
 $$0=\varepsilon_V^U\phi^\ell\phi_{;u_1}\mathcal{R}_{V,2}^{U,m+1}e^{v_1}.$$
 We set $u_1=k$ and then set $v_1=k$, $v_2=k$, ..., and $v_{m+1}=k$ in 
 turn to see:
 \begin{eqnarray*}
 0&=&\varepsilon_J^I\phi^\ell\phi_{;k}\mathcal{R}_{J,1}^{I,m}e^k
 -m\varepsilon_J^I\phi^\ell\phi_{;k}R_{i_1i_2j_2k}\mathcal{R}_{J,3}^{I,m}e^{j_1}\\
 &=&\Xi_{m+1,m}^{4,\ell}-m\Theta_{m+1,m}^{1,\ell}.
 \end{eqnarray*}
 We set $u_2=k$ and expand in $v$ to see:
 \begin{eqnarray*}
 0&=&\varepsilon_J^I\phi^\ell\phi_{;i_1}R_{i_2kj_2j_1}\mathcal{R}_{J,3}^{I,m}e^k
 -2\varepsilon_J^I\phi^\ell\phi_{;i_1}R_{i_2kj_2k}\mathcal{R}_{J,3}^{I,m}e^{j_1}\\
 &-&(m-2)\varepsilon_J^I\phi^\ell\phi_{;i_1}R_{i_2kj_3j_2}R_{i_3i_4j_4k}\mathcal{R}_{J,5}^{I,m}e^{j_1}\\
 &=&\Theta_{m+1,m}^{2,\ell}+2\Xi_{m+1,m}^{5,\ell}-(m-2)\Theta_{m+1,m}^{3,\ell}.
 \end{eqnarray*}
 Next, we set $v_1=k$ and expand in $u$ to see:
 \begin{eqnarray*}
 0&=&\varepsilon_J^I\phi^\ell\phi_{;k}\mathcal{R}_{J,1}^{I,m}e^k
 +m\varepsilon_J^I\phi^\ell\phi_{;i_1}R_{i_2kj_2j_1}\mathcal{R}_{J,3}^{I,m}e^k\\
 &=&\Xi_{m+1,m}^{4,\ell}+m\Theta_{m+1,m}^{2,\ell}.
 \end{eqnarray*} 
 Finally, we set $v_2=k$ and expand in $u$ to see:
 \begin{eqnarray*}
 0&=&\varepsilon_J^I\phi^\ell\phi_{;k}R_{i_1i_2j_2k}\mathcal{R}_{J,3}^{I,m}e^{j_1}
 +2\varepsilon_J^I\phi^\ell\phi_{;i_1}R_{i_2kj_2k}\mathcal{R}_{J,3}^{I,m}e^{j_1}\\
 &+&(m-2)\varepsilon_J^I\phi^\ell\phi_{;i_1}R_{i_2i_3j_2k}R_{i_4kj_4j_3}\mathcal{R}_{J,5}^{I,m}e^{j_1}\\
 &=&\Theta_{m+1,m}^{1,\ell}-2\Xi_{m+1,m}^{5,\ell}+(m-2)\Theta_{m+1,m}^{3,\ell}.
 \end{eqnarray*}
 We can show that
 $$
 \{\Theta^{1,\ell}_{m+1,m},\Theta^{2,\ell}_{m+1,m},\Theta_{m+1,m}^{3,\ell}\}\subset
 \Span\{\Xi_{m+1,m}^{i,\ell}\}_{i,\ell}
 $$
 by computing:
 \begin{eqnarray*}
 &&\Theta_{m+1,m}^{1,\ell}=\textstyle\frac1m\Xi_{m+1,m}^{4,\ell}=2\Xi_{m+1,m}^{5,\ell}-(m-2)\Theta_{m+1,m}^{3,\ell},\\
 &&\Theta_{m+1,m}^{2,\ell}=-\textstyle\frac1m\Xi_{m+1,m}^{4,\ell}=-2\Xi_{m+1,m}^{5,\ell}+(m-2)\Theta_{m+1,m}^{3,\ell}.
 \end{eqnarray*}
 Finally, we put $u_1=k$ in the identity
 $$0=\varepsilon_V^U\phi^{\ell} 
 R_{u_2u_3v_3v_2;u_1}\mathcal{R}_{V,4}^{U,m+1}e^{v_1}$$
 and expand in $v$ to show
 \begin{eqnarray*}
 0&=&\varepsilon_J^I \phi^\ell R_{i_1i_2j_2j_1;k} 
 \mathcal{R}_{J,3}^{I,m}e^k
 -2 \varepsilon_J^I \phi^\ell R_{i_1i_2j_2k;k} 
 \mathcal{R}_{J,3}^{I,m}e^{j_1}\\
 & & -(m-2) \varepsilon_J^I \phi^\ell R_{i_1i_2j_3j_2;k} R_{i_3i_4j_4k}
 \mathcal{R}_{J,5}^{I,m}e^{j_1} \\
 &=& \Xi_{m+1,m}^{2,\ell} -2 \Xi_{m+1,m}^{3,\ell} 
 +(m-2)\Theta_{m+1,m}^{4,\ell}
 \,.\end{eqnarray*}
 This establishes the lemma.
 \end{proof}

 We now prove Lemma \ref{lem1.3}. Let $m=2\bar m$ be even. We apply 
 Lemma \ref{lem3.3} and Lemma \ref{lem3.4} to see there
 exist universal constants so
 $$a_{m+2,m}^{d+\delta}(\phi,g)=\textstyle\sum_{i,\ell}c_{m+1,m}^{i,\ell}\delta_g\{\Xi_{m+1,m}^{i,\ell}\},$$
 where $\ell$ is chosen so $\phi$ appears an even number of times in 
 each expression.
 Terms which are linear in the $2$ jets of $\phi$ and which are of total 
 weight $2$ in $\phi$ can arise only from $i=4$
 and
 $i=5$. Consequently we have
 $$a_{m+2,m}^{d+\delta}(\phi,g)=-\textstyle\sum_\ell\phi^\ell\varepsilon_J^I\{
 c_{m+1,m}^{4,\ell}\phi_{;kk}\mathcal{R}_{J,1}^{I,m}
 +c_{m+1,m}^{5,\ell}\phi_{;i_1j_1}R_{i_2kkj_2}\mathcal{R}_{J,3}^{I,m}\}+\ldots
 $$
 where $\ell$ is odd. Replacing $\phi$ by $\phi+c$ does not change 
 $d_\phi$ and $\delta_{g,\phi}$. Thus
 $\phi^{\mu}$ does not appear in the formula for $a_{m+2,m}^{d+\delta}$ 
 for $\mu>0$. Consequently,
 \begin{equation}\label{eqn3.c}0=c_{m+1,m}^{4,\ell}\varepsilon_J^I\phi_{;kk}\mathcal{R}_{J,1}^{I,m}
 +c_{m+1,m}^{5,\ell}\varepsilon_J^I\phi_{;i_1j_1}R_{i_2kkj_2}\mathcal{R}_{J,3}^{I,m}.
 \end{equation}
 We consider the expressions:
 \begin{eqnarray*}
 &&A_1:=\phi_{;11}R_{1221}R_{3443}...R_{m-1,mm,m-1},\quad\text{and}\\
 &&A_2:=\phi_{;12}R_{1332}R_{3443}...R_{m-1,mm,m-1}.
 \end{eqnarray*}
 We may then expand
 \begin{eqnarray*}
 &&\phi_{;kk}\mathcal{R}_{J,1}^{I,m}\phantom{.............}=\phantom{a.........a}
 4^{\bar m}\bar m!A_1+\phantom{..............a........}0A_2+...,\\
 &&\phi_{;i_1j_1}R_{i_2kkj_2}\mathcal{R}_{J,3}^{I,m}=
 4^{\bar m-1}(\bar m-1)!A_1-4\cdot 4^{\bar m-1}(\bar m-1)!A_2+....
 \end{eqnarray*}
 Consequently equation 
 (\ref{eqn3.c}) implies $c_{m+1,m}^{4,\ell}=0$ 
 and $\quad c_{m+1,m}^{5,\ell}=0$.

 \smallbreak We argue similarly to show that if $\ell>0$, then
 \begin{eqnarray*}
 0&=&c_{m+1,m}^{1,\ell}\varepsilon_J^I\phi_{;j_2}\phi_{;i_1j_1}\phi_{;i_2}\mathcal{R}_{J,3}^{I,m},
 \quad\text{and}\\
 0&=&c_{m+1,m}^{2,\ell}\varepsilon_J^I\phi_{;k}R_{i_1i_2j_2j_1;k}\mathcal{R}_{J,3}^{I,m}
 +c_{m+1,m}^{3,\ell}\varepsilon_J^I\phi_{;j_2}R_{i_1i_2kj_1;k}\mathcal{R}_{J,3}^{I,m}.
 \end{eqnarray*}
 This shows $c_{m+1,m}^{1,\ell}=0$ for $\ell>0$. We consider the 
 expressions:
 \begin{eqnarray*}
 &&B_1:=\phi_{;1}R_{1221;1}R_{3443}...R_{m-1,mm,m-1},\quad\text{and }\\
 &&B_2:=\phi_{;3}R_{1221;3}R_{3443}...R_{m-1,mm,m-1}\end{eqnarray*}
 and expand
 \begin{eqnarray*}
 &&\varepsilon_J^I\phi_{;k}R_{i_1i_2j_2j_1;k}\mathcal{R}_{J,3}^{I,m}=\phantom{......a}
 4^{\bar m}(\bar m-1)!B_1+4^{\bar m}(\bar m-1)!B_2+\ldots,
 \\
 &&\varepsilon_J^I\phi_{;j_2}R_{i_1i_2kj_1;k}\mathcal{R}_{J,3}^{I,m}=
 2\cdot4^{\bar m-1}(\bar 
 m-1)!B_1+\phantom{..............a}0B_2+\ldots
 \end{eqnarray*}
 to see $c_{m+1,m}^{2,\ell}=0$ and $c_{m+1,m}^{3,\ell}=0$ for $\ell>0$; 
 Lemma \ref{lem1.3} now
 follows.
 \hfill\qedbox

 \section{Determining the normalizing constants}\label{Sect4}
 We complete the proof of Theorem \ref{thm1.2} by evaluating the
 normalizing constants of Lemma \ref{lem1.3}:

 \begin{lemma}\label{lem4.1} Let $m=2\bar m$. Then\begin{enumerate}
 \item $c_{m+2,m}^1=\frac{4\bar m}{\pi^{\bar m}8^{\bar m}\bar m!}$.
 \item $c_{m+2,m}^2=\frac1{12}\frac1{\pi^{\bar m}8^{\bar m}\bar m!}$.
 \item $c_{m+2,m}^3=\frac{\bar m}6\frac1{\pi^{\bar m}8^{\bar m}\bar 
 m!}$.
 \end{enumerate}
 \end{lemma}

 \begin{proof} We shall apply Theorem \ref{thm1.1}, Theorem \ref{thm2.1}, Lemma \ref{lem2.2}, and Lemma \ref{lem2.3}. We use the method of
universal examples. Give 
 $M:=S^{m-2}\times S^1\times S^1$ the product metric.
 Let $\phi=\phi_1(\theta_1)+\phi_2(\theta_2)$. Then:
 \begin{eqnarray*}
 &&a_{m+2,m}^{d+\delta}(\phi,g)=2c_{m+2,m}^12^{\bar 
 m-1}(m-2)!\phi_{;m-1m-1}\phi_{;mm}\\
 &&\quad=a_{m-2,m-2}^{d+\delta}(0,g_{S^{m-2}})\cdot 
 a_{2,1}^{d+\delta}(\phi_1,d\theta_1^2)\cdot
 a_{2,1}^{d+\delta}(\phi_2,d\theta_2^2)\\
 &&\quad=\textstyle\frac1{8^{\bar m-1}\pi^{\bar m-1}(\bar m-1)!}\cdot 
 2^{\bar m-1}(m-2)!
 \frac1\pi\phi_{1;m-1,m-1}\phi_{2;mm}.
 \end{eqnarray*}
 We solve this equation for $c_{m+2,m}^1$ to establish assertion (1).

 \smallbreak For the remainder of the proof of the Lemma, 
 we set $\phi=0$ to consider only
 metric invariants. We express
$$a_{m,m}^{d+\delta}=\mathcal{E}_{m,m}c_{m,m}\quad\text{for}\quad
\mathcal{E}_{m,m}=\varepsilon_J^I \mathcal{R}_{J,1}^{I,m}\quad\text{and}\quad
 c_{m,m}=\textstyle\frac1{8^{\bar m}\pi^{\bar m} \bar m!}.$$

If $m=2$, then the invariants 
 $(\varepsilon_J^I \mathcal{R}_{J,1}^{I,m})_{;kk}$ and 
 $(\varepsilon_J^I R_{i_1i_2kj_1;k}\mathcal{R}_{J,3}^{I,m})_{;j_2}$
 are not linearly independent.
If
 $(N,g_N)$ is a Riemann surface, then we may establish assertions (2) and (3) for $m=2$ by computing:
 \begin{eqnarray*}
 &&a_{2,2}^{d+\delta}(g_N)=\textstyle\frac1{4\pi}\sum_p(-1)^p\Tr(E^p)\\
 &&a_{4,2}^{d+\delta}(g_N)=\textstyle\frac1{4\pi}\frac16\{\sum_p(-1)^p\Tr(E^p)\}_{;kk}
 +O(R^2)\\
 &&\qquad=\textstyle\frac16\{a_{2,2}^{d+\delta}\}_{;kk}+O(R^2)
 =\textstyle\frac1{4\pi}\textstyle\frac16R_{ijji;kk}+O(R^2).\end{eqnarray*}

Suppose now that $m=4$. Since $\sum_p(-1)^p\Tr(\dim(\Lambda^p))=0$, we compute:
 \begin{eqnarray*}
&&0=a_{2,4}^{d+\delta}(0,g)=\textstyle\frac1{8\pi^2}\textstyle\sum_p(-1)^p\Tr(E^p),\\
 &&a_{4,4}^{d+\delta}=\textstyle\frac1{4^2\pi^2}\textstyle\sum_p(-1)^p
 \Tr\{\frac12E^pE^p+\frac1{12}\Tr(\Omega_{ij}^p\Omega_{ij}^p)\},\\
 &&a_{6,4}^{d+\delta}=\textstyle\frac1{4^2\pi^2}\textstyle\sum_p(-1)^p
 \Tr\{\frac1{45}\Omega_{ij;k}^p\Omega_{ij;k}^p+\frac1{180}\Omega_{ij;j}^p\Omega_{ik;k}^p
 +\frac{1}{60}\Omega_{ij;kk}^p\Omega_{ij}^p\\&&\textstyle\quad
 +\frac1{60}\Omega_{ij}^p\Omega_{ij;kk}^p+\frac1{6}E^pE_{;ii}^p
 +\frac1{12}E_{;i}^pE_{;i}^p\}+O(R^3).\end{eqnarray*}
 We study the expressions $C_1:=R_{1221}R_{3443}$ and 
 $C_2:=R_{1221;2}R_{3443;2}$ and suppress
 other terms. Only the term $E^pE^p$ can give rise to the expression $A_2$ 
 and only the term
 $E_{;i}^pE_{;i}^p$ can give rise to the expression $A_3$. We prove assertion (2) if $m=4$ by computing:
\medbreak\qquad
$a_{4,4}^{d+\delta}=
 \textstyle\frac1{2\cdot4^2\pi^2}\textstyle\sum_p(-1)^p\Tr(E^pE^p)+\ldots
 =\textstyle\frac{32}{8^2\pi^22!}R_{1221}R_{3443}+\ldots$
\smallbreak\qquad
$a_{6,4}^{d+\delta}=
 \textstyle\frac1{4^2\pi^2}\frac1{12}\textstyle\sum_p(-1)^p\Tr(E^p_{;i}E^p_{;i})+\ldots
 =\textstyle\frac1{4^2\pi^2}\frac1{24}\textstyle\sum_p(-1)^p\Tr(E^pE^p)_{;ii}+\ldots$
\smallbreak\qquad\qquad\ 
$=\textstyle\frac1{12}(a_{4,4}^{d+\delta})_{;kk}+\ldots
 =\textstyle\frac1{6}\textstyle\frac{32}{8^2\pi^22!}R_{1221;2}R_{3443;2}+\ldots$
\smallbreak\qquad\qquad\ 
$=2c_{6,4}^2\varepsilon_J^I
 R_{i_1i_2j_2j_1;k}R_{i_3i_4j_4j_3;k}+\ldots=
 64c_{6,4}^2R_{1221;2}R_{3443;2}+\ldots,\quad\text{so}$
\smallbreak\qquad
$c_{6,4}^2=\textstyle\frac1{12}\frac1{8^2\pi^22!}$.
\medbreak\noindent
If $m>4$, let
 $(M,g):=(N^4\times S^{m-4},g_N+g_0)$. Assertion (2) follows in general from:
\medbreak\quad\ \ 
$a_{m+2,m}^{d+\delta}(g)
=\bar m(\bar m-1)c_{m+2,m}^2\varepsilon_J^I
 (R_{i_1i_2j_2j_1;k}R_{i_3i_4j_4j_3;k})(g_N)\mathcal{E}_{m-4,m-4}(g_0)
 +\ldots$
\smallbreak\qquad\qquad$=a_{6,4}^{d+\delta}(g_N)a_{m-4,m-4}^{d+\delta}(g_0)
 +\ldots$
\smallbreak\qquad\qquad$=2c_{6,4}^2c_{m-4,m-4}\varepsilon_J^I
 (R_{i_1i_2j_2j_1;k}R_{i_3i_4j_4j_3;k})(g_N)\mathcal{E}_{m-4,m-4}(g_0)
 +\ldots\quad\text{so}$
\smallbreak\quad\ \ $c_{m+2,m}^2=\textstyle\frac2{\bar m(\bar 
 m-1)}\frac1{12}\frac{1}{\pi^28^22!}
 \cdot\frac1{\pi^{\bar m-2}8^{\bar m-2}(\bar m-2)!}
 =\textstyle\frac1{12}\frac1{\pi^{\bar m}8^{\bar m}\bar m!}$.

\medbreak Let $(M,g)=(N^2\times
 S^{m-2},g_N+g_0)$. We derive a
 relation between the invariants $c_{m+2,m}^2$ and $c_{m+2,m}^3$
 to complete the proof of assertion (3):
 \medbreak\qquad
 $a_{m+2,m}^{d+\delta}(g)
 =\textstyle(2\bar 
 mc_{m+2,m}^2+c_{m+2,m}^3)R_{ijji;kk}(g_N)\cdot\mathcal{E}_{m-2,m-2}(g_0)$
 \medbreak\qquad$\quad=a_{4,2}^{d+\delta}(g_N)a_{m-2,m-2}^{d+\delta}(g_0)=\textstyle\frac1{4\pi}\frac16R_{ijji;kk}(g_N)\cdot
 c_{m-2,m-2}\mathcal{E}_{m-2,m-2}(g_0)$\quad so
 \medbreak\qquad$\textstyle 2\bar
 mc_{m+2,m}^2+c_{m+2,m}^3=\frac16\frac1{4\pi}\frac1{\pi^{\bar 
 m-1}8^{\bar m-1}(\bar m-1)!} =\frac{\bar
 m}3\frac1{\pi^{\bar m}8^{\bar m}\bar m!}.$
 \end{proof}

 \section*{Acknowledgments} Research of PG partially supported by the 
 NSF (USA), the MPI (Leipzig, Germany), and the
 Mittag-Leffler (Stockholm, Sweden). 
 Research of KK and DV supported by the MPI 
 (Leipzig, Germany).


\begin{thebibliography}{AAA}

 \smallbreak\bibitem{AmBeOc89} P. Amsterdamski, A. Berkin, and D. 
 O'Connor,
 {\it $ b \sb{ 8}$ Hamidew coefficient for a scalar field},
 Classical Quantum Grav.,
 {\bf 6} (1989), 1981--1991.

 \bibitem{ABI}
 A. A. Andrianov, N. V. Borisov and M. V. Ioffe,
 {\it Factorization Method And Darboux Transformation 
 For Multidimensional Hamiltonians,}
 Theor.\ Math.\ Phys.\ {\bf 61} (1984) 1078
 [Teor.\ Mat.\ Fiz.\ {\bf 61} (1984) 183].

 \smallbreak\bibitem{Av90} I. G. Avramidi, 
 {\it The covariant technique for the calculation of the heat kernel 
 asymptotic
 expansion},
 Physics Letters B.,
 {\bf 238} (1990), 92--97.

 \smallbreak\bibitem{ABP73} M. F. Atiyah, R. H. Bott, and V. K. Patodi,
 {\it On the heat equation and the index theorem},
 Invent. Math. {\bf 19} (1973), 279--330; Errata {\bf 28} (1975), 
 277--280.

 \smallbreak\bibitem{C44} S. Chern,
 A simple intrinsic proof of the Gauss-Bonnet formula for closed 
 Riemannian manifolds,
 Ann. of Math. 
 \bf 45 \rm(1944), 741--752.

 \smallbreak\bibitem{PG73} P. Gilkey, 
 {\it Curvature and the eigenvalues of the Laplacian for elliptic 
 complexes},
 Advances in Math, 
 {\bf 10} (1973), 344--382.

 \smallbreak\bibitem{PG75} ---,
 Local invariants of an embedded Riemannian manifold,
 {\it Annals of Math.}, {\bf 102} (1971), 187--203.

 \smallbreak\bibitem{PG79} ---,
 Curvature and the heat equation for the DeRham
 complex, in
 {\bf Geometry and Analysis (Papers dedicated to the
 memory of V.K.Patodi)},
 Indian Academy of Sciences (1979), 47--80.

 \smallbreak\bibitem{PG94} ---,
 {\bf Invariance Theory, the
 heat equation, and the Atiyah-Singer index theorem} $2^{nd}$ ed., CRC 
 Press
 ISBN 0-8493-7874-4 (1994), 516pp.

 \bibitem{GKV02a} P. Gilkey, K. Kirsten, and D. Vassilevich, {\it 
 Divergence terms in the supertrace heat
 asymptotics for the de Rham complex on a manifold with boundary}, 
 math-ph/0211020.

 \bibitem{GKVZ02} P. Gilkey, K. Kirsten, D. Vassilevich, and A. 
 Zelnikov, {\it Duality symmetry of the
 $p$-form effective action and super trace of the twisted de Rham
 complex}, hep-th/0209125, to appear in Nucl. Phys. B.

 \smallbreak\bibitem{McSi67} H. P. McKean and I. M. Singer, 
 {\it Curvature and the eigenvalues of the Laplacian}, 
 J. Diff. Geo.,
 {\bf 1} (1967), 43--69.

 \smallbreak\bibitem{Pa70} V. K. Patodi,
 Curvature and the fundamental solution of the heat operator,
 J. Indian Math. Soc.
 {\bf 34} (1970), 269--285.

 \smallbreak\bibitem{Pol00} I. Polterovich, Heat invariants of 
 Riemannian manifolds,
 {\it Isr. J. Math.} {\bf 119} (2000), 239-252.


 \smallbreak\bibitem{Se68} R. Seeley,
 {\it Complex powers of an elliptic operator}, 
 in {\bf Amer. Math. Soc. Proc. Symp. Pure Math},
 {\bf 10} (1968), 288--307.

 \smallbreak\bibitem{Ven98} A. E. van de Ven,
 {\it Index-free heat kernel coefficients},
 Class. Quant. Grav.,
 {\bf 15} (1998), 2311--2344,
 hep-th/9708152.

 \smallbreak\bibitem{VZ00} D. Vassilevich and A. Zelnikov,
 {\it Discrete symmetries of functional determinants},
 Nucl. Phys. B {\bf 594} (2000), 501--517.

 \smallbreak\bibitem{We46} H. Weyl,
 {\bf The Classical Groups},
 Princeton Univ. Press, Princeton, 1946.

 \bibitem{Wit82} E. Witten, {\it Supersymmetry and Morse Theory}, J. 
 Diff. Geom. {\bf 17} (1982), 661--692.
 \end{thebibliography}
 \end{document}